\documentclass[aps,prb,twocolumn,showpacs]{revtex4-1}

\usepackage{graphicx}
\DeclareGraphicsExtensions{.png}
\DeclareGraphicsExtensions{.jpeg}

\usepackage{xcolor}
\usepackage{amsmath}
\usepackage{amssymb}

\usepackage{dcolumn}% Align table columns on decimal point
\usepackage{bm}% bold math
\usepackage{hyperref}
\usepackage[latin1]{inputenc}
\usepackage[american]{babel}
\usepackage{latexsym}
\usepackage{float}

\begin{document}

\title{Topological $\pi$-junctions from crossed Andreev reflection in the Quantum Hall regime}

\index{F. Finocchiaro}
\index{F. Guinea}
\index{P. San-Jose}

\author{F. Finocchiaro$^{1,2}$, F. Guinea$^{2,3}$ and P. San-Jose$^{1}$}
\affiliation{${}^1$Materials Science Factory, ICMM-CSIC, Sor Juana Ines de La Cruz 3, 28049 Madrid, Spain}
\affiliation{${}^2$IMDEA Nanociencia, Calle de Faraday 9, 28049 Madrid, Spain}
\affiliation{${}^3$Department of Physics and Astronomy, University of Manchester, Manchester M13 9PL, United Kingdom}

\date{\today}

%%%%%%%%%%%%%%%%%%%%%%%%%%%%%%%%%%%%%%%%%%%%%%%%%%%%%%%%%%%%%%%%%%%%%%%%%%%%%%%%

\begin{abstract}
We consider a two-dimensional electron gas (2DEG) in the Quantum Hall regime in the presence of a Zeeman field, with the Fermi level tuned to filling factor $\nu=1$. We show that, in the presence of spin-orbit coupling, contacting the 2DEG to a narrow strip of an s-wave superconductor produces a topological superconducting gap along the contact as a result of crossed Andreev reflection (CAR) processes across the strip. The sign of the topological gap, controlled by the CAR amplitude, depends periodically on the Fermi wavelength and strip width and can be externally tuned. An interface between two halves of a long strip with topological gaps of opposite sign implements a robust $\pi$-junction, hosting a pair of Majorana zero modes that do not split despite their overlap. We show that such a configuration can be exploited to perform protected non-Abelian tunnel-braid operations without any fine tuning.
\end{abstract}

\maketitle

%%%%%%%%%%%%%%%%%%%%%%%%%%%%%%%%%%%%%%%%%%%%%%%%%%%%%%%%%%%%%%%%%%%%%%%%%%%%%%%%

%\section{Introduction}

% Genric introduction to MZMs
During the last decade we have witnessed a surge in both theoretical and experimental progress towards the realisation of Majorana-based quantum computation.\citep{Alicea:RPP12,Deng:NL12,Williams:PRL12,Mourik:Science12,
Rokhinson:NatPhys12,Das:NatPhys12,Finck:PRL13,Nadj-Perge:Science14,Ruby:PRL15,Albrecht:Nature16,Deng:Science16,Suominen:PRB17,Suominen:17} Majorana zero modes (MZMs) are zero-energy bound quasiparticles of topological origin that are their own self-adjoint and obey non-Abelian anyon statistics. As a result, the adiabatic exchange (or `braiding') of a pair of MZMs rotates the wavefunction of the degenerate ground state in a non-commutative fashion.\citep{Ivanov:PRL01,Alicea:Nature11,Clarke:PRB11,Sau:PRB11,Halperin:PRB12,
Hyart:PRB13} Such a process or its generalisations\citep{Bonderson:PRL08,vanHeck:NJP12,Bonderson:PRB13,Vijay:PRB16,
Karzig:PRB17,Gharavi:PRB16,Hoffman:PRB16,Aasen:PRX16} can be viewed as a coherent manipulation of qubit states realised by pairs of MZMs. The interest in Majorana-based topological quantum computation stems from the fact that, as a result from the non-locality of the MZMs, local sources of noise do not
%, in principle,
affect the fidelity of the braiding operation, nor do they induce decoherence of the ground state manifold. This property has inspired implementations of fault-tolerant computation schemes able in principle to beat decoherence at the hardware level.\citep{Alicea:RPP12}

The fundamental ingredient needed to create MZMs is topological superconductivity, either intrinsic, like in p-wave superconductors,\citep{Kitaev:PU01,He:Science17} or artificially designed, like in proximitised superconducting wires with strong spin-orbit coupling (SOC) in an external magnetic field.\citep{Oreg:PRL10,Lutchyn:PRL10,Sau:PRL10}
% 2DEGs and our reference experiment
More recently, two dimensional electron gases (2DEGs) with induced superconductivity are being actively investigated as platforms for topological superconductivity.\citep{Takayanagi:PRL95,Bauch:PRB05,SanJose:PRX15,Shabani:PRB16,Kjaergaard:NatComms16,Hell:PRL17,
Pientka:PRX17,Hell:PRB17,Nichele:17,Suominen:17,Suominen:PRB17}
In addition to the increased freedom afforded by the planar geometry, these systems allow for the formation of a new type of topological quasi-one dimensional (1D) system, confined on both sides by two different superconductors with a phase difference $\pi$.  For transparent enough contacts, such $\pi$ junctions can greatly reduce the magnetic fields required for MZMs to emerge.\citep{Hell:PRL17,Pientka:PRX17}
%Hence, creating and controlling $\pi$ junctions has recently become a powerful tool to facilitate the formation of topological superconductivity and MZMs in planar systems.

\begin{figure}
\centering
\includegraphics[width=8.4cm]{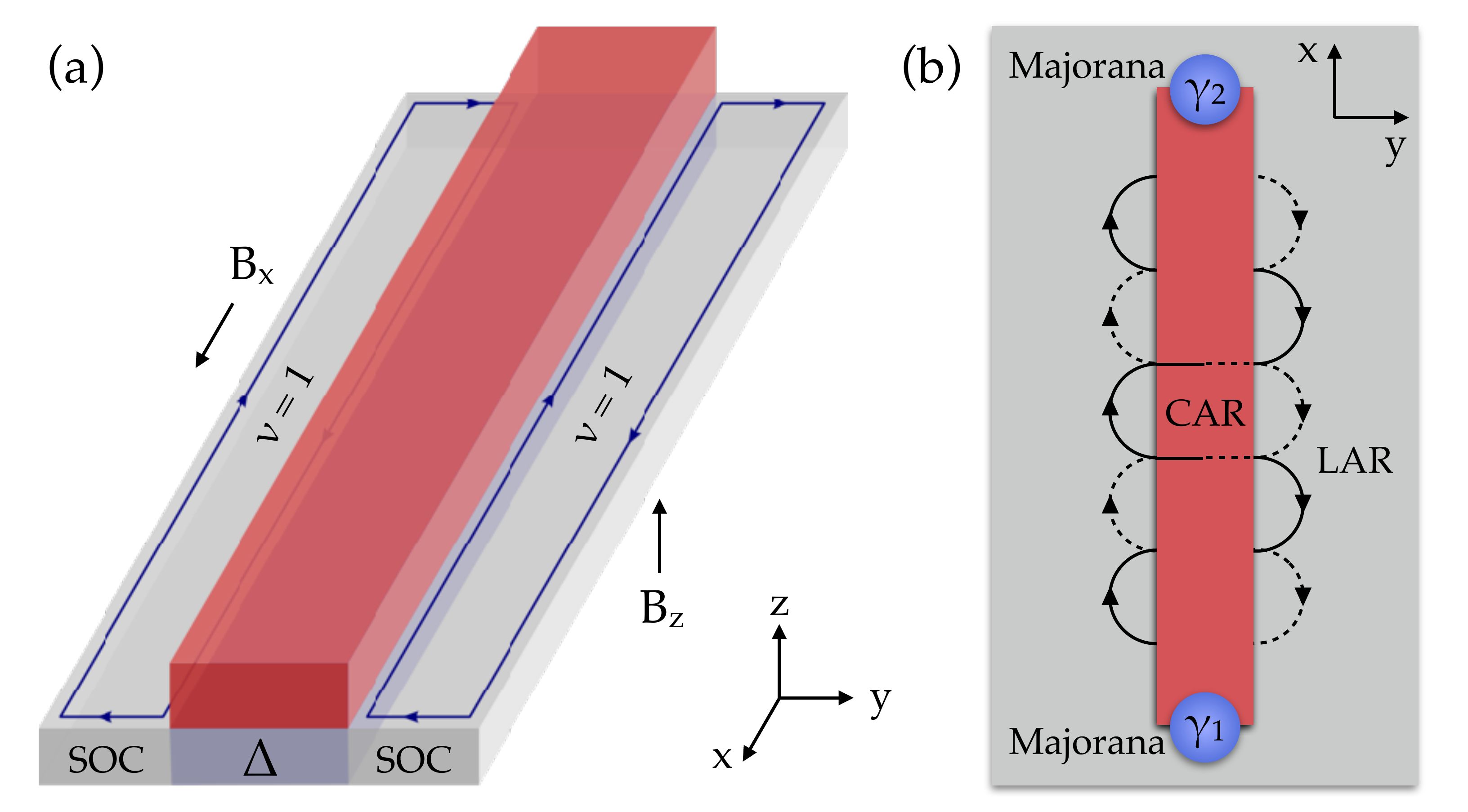}
\caption{(a) A 2DEG with strong SOC and in the $\nu=1$ state of the QH phase is proximized along a narrow strip with an s-wave superconductor. (b) Sketch of the crossed Andreev reflection (CAR) and local Andreev reflection (LAR) processes occurring across and along the proximized region, respectively. Full (dashed) lines represent electrons (holes). CAR processes induce a topological gap in the edge states Majorana zero modes at the ends of the strip.}
\label{fig:fig1}
\end{figure}

% Our idea: what we do
In this work we show that planar junctions allow for yet another implementation of 1D topological superconductivity, with a geometry dual to the above. It is achieved by contacting a long and narrow strip of a conventional superconductor to a 2DEG in the Quantum Hall (QH) regime at filling factor $\nu=1$. The proximitised region acquires a superconducting gap $\Delta$, and as a results develops gapless QH edge states along each side. Due to local Andreev reflection (LAR) processes, these edge states are a mixture of electrons and holes,\cite{Hoppe:PRL00,Qi:PRB10} see Fig. \ref{fig:fig1}. Assuming that spin-orbit coupling (SOC) is present in the system, and that the strip width is comparable with or smaller than the superconducting coherence length, the QH edge states may become Cooper-paired through additional crossed Andreev reflection (CAR) processes\citep{Byers:PRL95,denHartog:PRL96,Deutscher:APL00,Yeyati:NatPhys07,Lee:NatPhys17} across the strip. We show that a topologically non-trivial superconducting gap $\Delta^*$ then opens in the edge state dispersion, and MZMs emerge at either end of the strip. This possibility was suggested by Lee \emph{et al.} in Ref. \onlinecite{Lee:NatPhys17}, where the requisite CAR processes were experimentally demonstrated in the case of graphene, although they concentrated on the $\nu=2$ regime and not on the $\nu=1$ condition required for the formation of MZMs.
%\redmark{Related approaches for engineering parafermions through CAR in fractionalized QH phases have also been explored\cite{Clarke:Nature13,Klinovaja:PRB14,Alicea:AR16}.}

Here we theoretically investigate the conditions for CAR-induced topological superconductivity at $\nu=1$. \footnote{We note that for $\nu=2$ \cite{Lee:NatPhys17} (or even fillings in general), pairs of Majoranas will be generated which will not be protected against hybridisation into conventional fermions. Odd fillings, however, will always generate one protected unpaired Majorana zero mode.}
(Related approaches have been explored in fractionalized QH systems supporting parafermions\cite{Clarke:Nature13,Klinovaja:PRB14,Alicea:AR16}.)
We find that both the magnitude and, more importantly, the \emph{sign} of the topological gap depends on the amplitude of the CAR processes. As a result, the sign of $\Delta^*$ can be controlled by adjusting the width of the strip and/or the electronic density of the proximitised region, which in turn determine the CAR amplitude. Reeg \emph{et al.} anticipated such a possibility while studying a related system of two parallel nanowires coupled through a superconductor.\citep{Reeg:PRB17}
We show that this effect may be used to induce a sign change $\Delta^*\to -\Delta^*$ along the strip by e.g. electrostatic gating. This situation corresponds to a one-dimensional topological $\pi$-junction along the strip which is host to two degenerate MZMs that do not hybridise despite their spatial overlap.\cite{Ojanen:PRB13,Klinovaja:PRB14, Schrade:PRL15}
Since the original induced $\Delta$ does not change sign (only the edge state gap $\Delta^*$ does), no external fine-tuning is required to mantain the $\pi$ phase difference, and the MZMs remain protected at zero energy. As we will show, this allows for a powerful generalisation of tunnel-braiding strategies (originally proposed by Flensberg \citep{Flensberg:PRL11}) on the two MZMs in the junction, without the need to carefully control external parameters in the process.

Consider a normal (N) 2DEG  with a proximitised superconducting strip (S) of width $W_S$ along the $x$ direction, see Fig. \ref{fig:fig1}a. The N region is in the QH regime and is subject to a Zeeman field along $x$ allowing the electron density to be tuned to an odd filling factor $\nu=1$. (Other mechanisms such as interaction-induced spin instabilities may play the role of the Zeeman field in some systems\citep{Lado:PRB14,Finocchiaro:2M17}). The S region has uniform superconducting pairing $\Delta$ induced by proximity to the parent superconductor. We also assume that SOC is present in the system, either in the N region and/or in the S region (e.g. inherited from a superconductor made of heavy elements, such as NbN or NbTiN). The electronic structure of this system, obtained using a tight-binding approximation on a square lattice (see Supplementary Information\cite{SupInfo} for details), is studied in the following.

\begin{figure}
\centering
\includegraphics[width=8.6cm]{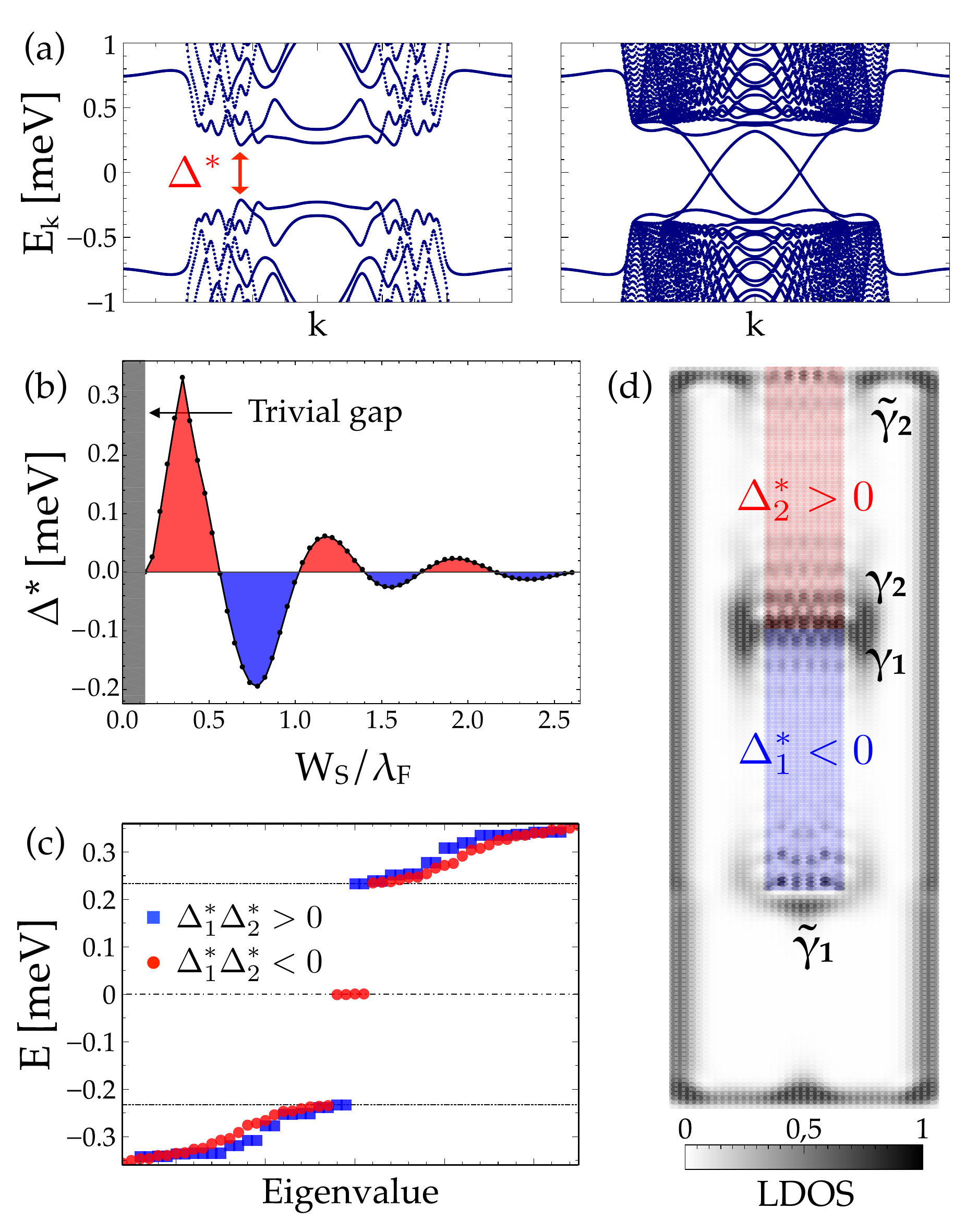}
\caption{(a) Spectrum of the system with periodic boundary conditions (PBC) along both directions for widths of the central strip such that CAR is present (left, $W_S=300$ nm) and absent (right, $W_S=2\,\mu$m), respectively. (b) Behavior of the topological gap $\Delta^*$ as a function of $W_S/\lambda_F$, for $\mu/\Delta=1.95$. The grey region corresponds to the opening of a trivial gap due to the direct overlap of the QH edge states. (c) Lowest eigenvalues in a system with PBC and two gaps $\Delta^*_{1,2}$ along the strip, either of equal (blue) or opposite sign (red). Two pairs of MZMs appear in the latter case (one pair at each of the two junctions, required in the case of PBC) (d) LDOS associated to the zero energy eigenvalues, calculated for a system with open boundaries. The strip is such that it terminates on one end within the 2DEG and on the other at the sample edge. One Majorana ($\tilde{\gamma}_1$) is therefore localized at one end and the other ($\tilde{\gamma}_2$) delocalises in the QH edge states. The gap $\Delta^*$ changes sign along the strip length so that two additional non-hybridising localized MZMs $\gamma_1$ and $\gamma_2$ appear at the boundary.
See Supplementary Information\cite{SupInfo} for parameters used.}
\label{fig:fig2}
\end{figure}

% Topological gap
Since the N region is in the $\nu=1$ QH regime and the S strip is trivially gapped, each of the two NS interfaces hosts a single spin-polarized edge state. These states travel in opposite directions at opposite interfaces (see Fig. \ref{fig:fig1}a).
Local Andreev reflections at each interface transform the edge states into coherent superpositions of electrons and holes,\citep{Hoppe:PRL00,Qi:PRB10, Chung:PRB11, Wang:PRB15, He:Science17} but do not open a gap because of the chiral nature of the carriers.
%, known as chiral Majoranas,\citep{Qi:PRB10, Chung:PRB11, Wang:PRB15, He:Science17} propagating along the interface in the same direction but with opposite wavevector (see Fig. \ref{fig:fig1}b).
%Because of the chirality of carriers, LAR processes do not open a gap in the QH edge states.
However, in our geometry with two parallel NS interfaces at either side of the strip, another type of Andreev reflection process can take place, wherein an electron on one interface is scattered as a hole into the \emph{other} interface.
This crossed Andreev reflection process has a significant amplitude only for strips narrower than the coherence length $\xi\approx \hbar v_F/\Delta$. Unlike local Andreev reflection, CAR processes may open a superconducting gap $\Delta^*$ in the presence of SOC, since electron and hole edge states at opposite interfaces propagate in opposite directions at the same wave vector. 
The role of the SOC is to cant the spin away from the Zeeman field in opposite directions in the two edge states, so that they can pair to form a spin singlet.
At $\nu=1$ the gap resulting from CAR is topological, as can be seen by a direct mapping of the two spin-canted edge states plus pairing into an Oreg-Lutchyn model\citep{Oreg:PRL10, Lutchyn:PRL10} [Eq. (B4) in Supplementary Information].
Fig. \ref{fig:fig2}a shows the gapped bandstructure of an infinite strip with significant CAR processes (left, $W_S \simeq \xi$) and the gapless case without CAR (right, $W_S \gg \xi$). 
The topological nature of $\Delta^*$ manifests in the appearance of MZMs when the strip is terminated inside the 2DEG (Fig. \ref{fig:fig2}d). 
%These MZMs can be either localised or delocalised depending on the type of termination. If the end of the strip lies inside the 2DEG, a localised Majorana bound state emerges at the boundary between the proximitised strip and the QH gas. However, if the strip ends close to the edge of the 2DEG, the MZM then hybridizes with the QH edge states and becomes delocalised into the continuum away from the end of the strip. This is illustrated in Fig. \ref{fig:fig2}d.

%\paragraph*{Topological gap.--}
The value of the topological gap $\Delta^*$ is entirely determined by the CAR amplitude, that in turn depends on the strip width $W_S$, the Fermi wavelength $\lambda_F$ and the singlet amplitude governed by the proximity gap $\Delta$ and the SOC strength $\alpha$. 
%which is itself a measure of the wavefunction overlap of opposite edge states under the strip. The overlap is determined by the interplay between the strip width $W_S$, the proximity gap $\Delta$ under the strip and the Fermi wavelength $\lambda_F$, that in turn depends on the SOC strength $\alpha$ and the chemical potential $\mu$. As a result, the topological gap $\Delta^*$ depends on all these parameters.
We have performed tight-binding simulations which show, specifically, that $\Delta^*$ is a real periodic function of the $W_S/\lambda_F$ with alternating sign, see Fig. \ref{fig:fig2}b. This behavior is confirmed by an analytical calculation in terms of Green's functions, which yields
%\begin{equation}
%\Delta^*\approx i\chi W_S\sqrt{\frac{m^3}{2}}
%\sum_{\eta=\pm}
%\eta\sqrt{\tilde{\mu}_{\eta}}
%      \csc\left(\sqrt{2mW_S^2\tilde{\mu}_\eta}\right)
%\end{equation}
\begin{equation}
\Delta^*\approx\frac{4\pi^2{t'}^2 a^3}{W_S\tilde\lambda_F^2 \tilde\mu}\times\,\mathrm{Im}\left(z\csc z\right)\sin\theta
\end{equation}
where $\theta$ is the spin canting angle due to spin-orbit coupling, $\tilde\lambda_F=2\pi/\sqrt{2m \tilde\mu}$, $\tilde\mu=\mu-k_F^2/2m$, $\mu$ is the strip Fermi energy, $k_F$ is the edge-state Fermi wavevector, and $z=2\pi\sqrt{1+i\Delta/\tilde\mu}\times W_{S}/\tilde\lambda_F$ (See Supplementary Information\cite{SupInfo} for details). This formalises the central finding of our work. The sign of $\Delta^*$ follows the change in the number of normal modes in the strip, given by $\lfloor 2W_{S}/\tilde\lambda_F\rfloor$. It is therefore likely to be realistically tuneable with electrostatic gating of the strip region that may modify both its effective width $W_S$ and electronic density, or by adjusting the width lithographically.\cite{Reeg:PRB17}

The possibility of changing the sign of the topological gap
%at a certain point
along the strip opens a new opportunity for the generation of MZMs. A long strip with uniform induced gap $\Delta$ but edge-state gaps of opposite sign in its two halves ($\Delta_1^*\Delta_2^*<0$) forms a topological $\pi$-junction, similar to a topological Josephson junction tuned to phase difference $\phi=\pi$. Such a system then develops two MZMs localised at the junction [see Fig. \ref{fig:fig2}(c,d)], that stay at zero energy despite their spatial overlap as long as the phase difference across the junction remains $\pi$. The $\pi$ phase difference between the two halves of the strip is robust. Since $\Delta^*$ on both sides is finite and real, its sign does not depend on perturbations. The CAR $\pi$-junction is furthermore stabilised by the phase rigidity of the strip order parameter $\Delta$.\cite{SupInfo} Unlike in $\phi=\pi$ Josephson junctions, it does not require fine tuning of any external parameter such as the flux across the superconducting circuit or the strip parameters.  As a result, CAR-induced topological superconductivity enables the creation of MZMs that remain decoupled regardless of their overlap. This offers great advantages in the context of coherent Majorana qubit manipulation and braiding, as outlined in the following.

%\paragraph*{\redmark{Tunnel-braid operations.---}}
We now present a possible application of the CAR $\pi$-junction to the challenge of non-Abelian Majorana braiding.
Plenty of proposals for the demonstration of the non-Abelian statistics of Majorana excitations have been presented which hinge on the physical exchange (or braiding) in real space of pairs of Majoranas.\citep{Ivanov:PRL01,Alicea:Nature11,Sau:PRL10,Sau:PRB11,Halperin:PRB12} Some other approaches, however, rest upon schemes that involve rotation of the wavefunction without the need for actual MZMs to move spatially.\citep{Bonderson:PRL08,Clarke:PRB11,Flensberg:PRL11,vanHeck:NJP12,Hyart:PRB13,
Bonderson:PRB13,Vijay:PRB16,Karzig:PRB17} Among these, it has been suggested\citep{Flensberg:PRL11} that adiabatic tunnel processes of single electrons from a quantum dot into pairs of Majorana zero modes can result in arbitrary non-Abelian rotations of the ground-state manifold.
%These so-called \textit{tunnel-braid} operations overperform braiding in that they allow to generate a universal set of single-qubit gates.
%These so-called \textit{tunnel-braid} operations extend braiding in that they allow to generate a universal set of single-qubit gates.
These so-called \textit{tunnel-braid} operations are extremely versatile as they allow a universal set of single-qubit gates, in contrast to braiding that only allows a limited set of operations.
%, whereas standard braiding can merely rotate the qubit around the $z$ axis by an angle $\pi/2$.
Unfortunately, tunnel-braiding has the drawback of requiring a precise, typically fine-tuned, phase difference of $\pi$ between the MZMs involved throughout the operation.
If the phase deviates from this value, the result of the operation becomes time-dependent and is no longer protected against decoherence. 
\begin{figure}
\centering
\includegraphics[width=8.7cm]{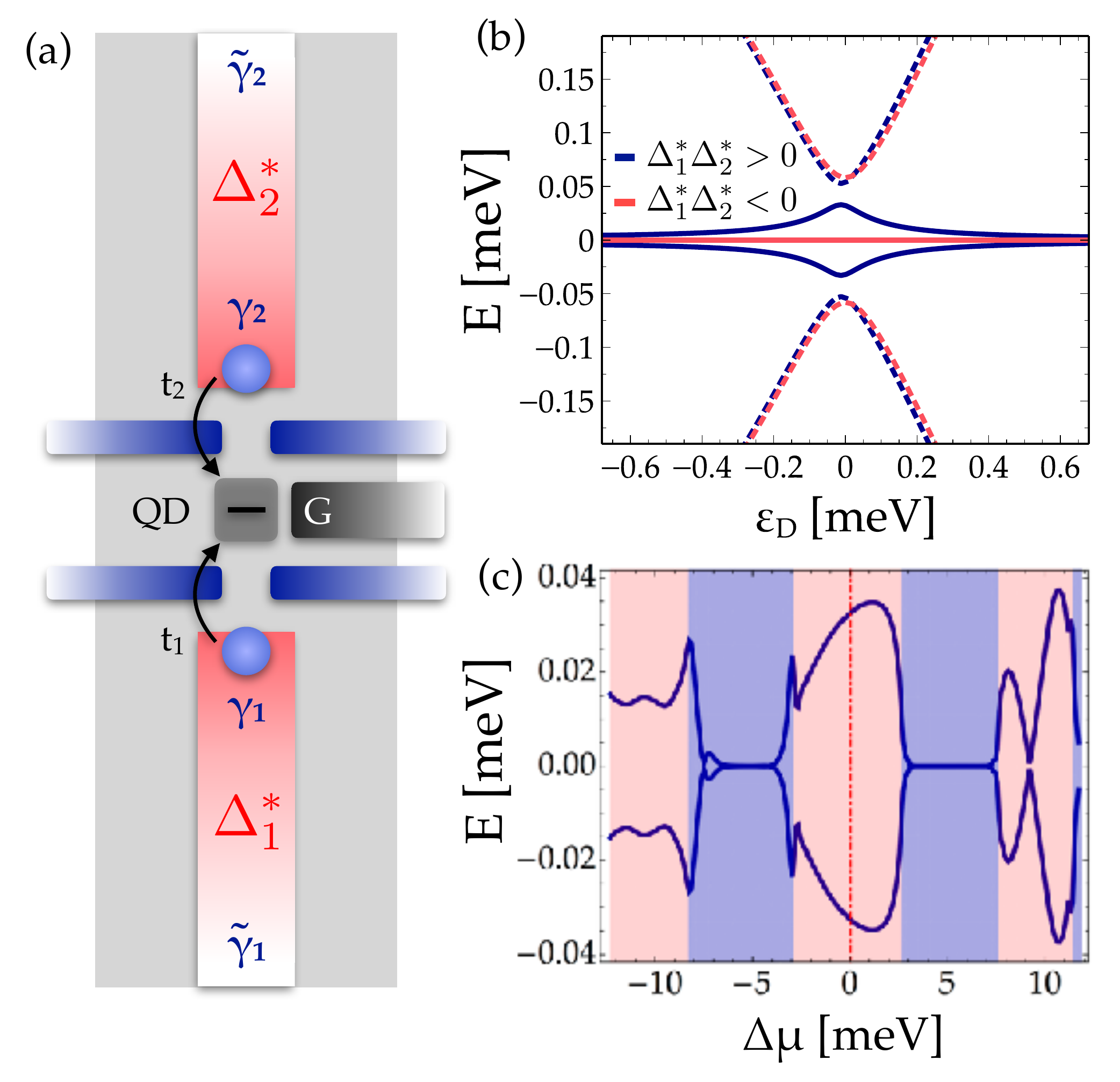}
\caption{(a) Sketch of a tunnel-braiding setup, with the two inner MZMs $\gamma_1$ and $\gamma_2$ from strips 1 and 2 coupled to a dot (QD) in the Coulomb blockade regime through tunnel barriers. The dot occupancy is controlled by a gate (G), which shifts the dot level $\varepsilon_D$. (b) Comparison of the low energy spectra  of the composite system for gaps $\Delta_{1,2}^*$ in the two strips of equal (blue) and opposite (red) sign.  The dotted and solid lines correspond predominantly to dot and Majorana states, respectively. (c) Energy of the MZMs as they hybridize through the dot as a function of the changing chemical potential between the left and right strips $\Delta\mu=\mu_1-\mu_2$, for $\mu_2$ and $\varepsilon_D$ fixed. The blue (dark) regions indicate phases where $\Delta^*_{1,2}$ have opposite sign, and $\gamma_{1,2}$ do not hybridise through the dot. See Supplementary Information\cite{SupInfo} for parameters used.}
\label{fig:fig3}
\end{figure}

The robustness and lack of fine-tuning of CAR $\pi$-junctions promises to overcome this problem. 
In Fig. \ref{fig:fig3}a we present a possible geometry to implement a CAR-protected tunnel-braiding scheme. We deposit two narrow superconducting strips on a $\nu=1$ 2DEG such that two independent CAR-induced topological gaps $\Delta_1^*$ and $\Delta_2^*$ open on each. One end of each strip terminates inside the 2DEG, so that the corresponding MZMs $\gamma_{1,2}$ lie within a finite distance of each other. The MZMs $\tilde{\gamma}_{1,2}$ on far end of the strips are assumed sufficiently far from the junction so as to become decoupled from $\gamma_{1,2}$. We control the Fermi level of the two strips, $\mu_1$ and $\mu_2$, by means of two independent gates, in order to tune the magnitude and sign of the topological gaps $\Delta^*_{1,2}$. 

The two `inner' MZMs $\gamma_1$ and $\gamma_2$ are then coupled to a quantum dot through two tunnel barriers that may be tuned externally. The tunnelling couplings $t_{1,2}$ control the specific non-Abelian opearation to perform. The dot is in the Coulomb-blockade regime, with occupation $N$. We adiabatically tune the dot level $\varepsilon_D$ across an $N\to N-1$ transitions between two adjacent Coulomb valleys. This transfers a single electron to the composite state of the two Majorana modes. Figure \ref{fig:fig3}b shows the evolution of the low-energy single-particle Bogoliubov spectrum of the full dot-2DEG-strip system across this process, with dashed lines corresponding to mostly-dot states, and solid lines to MZMs states in the strip. The two cases with equal (blue, $\phi=0$) and opposite (red, $\phi=\pi$) signs for $\Delta^*_{1,2}$ show markedly different structure. The conventional $\phi=0$ case splits the MZMs away from zero close to the $N\to N-1$ transition, as they become resonantly coupled via the dot state.\cite{Prada:PRB17} Such an operation is not protected against noise and its result depends on timing. In contrast, the $\phi=\pi$ case shows MZMs that remain exactly at zero energy throughout the operation, as their hybridisation across the dot is forbidden by the opposite sign of $\Delta^*_{1,2}$. The state after emptying the dot is then independent of timing and insensitive to noise in $\varepsilon_D$. As shown by Flensberg,\citep{Flensberg:PRL11} the transformation $P$ within the degenerate ground state manifold associated to this process is a rotation by an angle $\pi$ around an axis in the $xy$ plane, controlled by the tunnel couplings $t_{1,2}$.
If the couplings are then changed to $t'_{1,2}$, and the reverse adiabatic transition $N - 1 \to N$ on the dot is performed, the composite operation $P'P$ rotates the quantum state of the Majoranas by an arbitrary angle around the $z$ axis.
In comparison, braiding two MZMs can only rotate the wavefunction about the $z$ axis by an angle of $\pi / 2$.
%Remarkably, upon acting with composite $P'P$ operations on the fixed parity subspace of four MZMs coupled to a dot, any arbitrary SU(2) rotation can be achieved.

As no fine-tuning is required to maintain the $\phi=\pi$ condition in the CAR $\pi$-junction, the tunnel-braiding process should enjoy similar topological protection as a standard spatial-braiding.
In Fig. \ref{fig:fig3}c we show the MZM splitting across a resonant dot as we vary the Fermi energy under one of the strips, while the other is kept fixed. As expected, we find alternating $\phi=0$ (red) and $\phi=\pi$ (blue) regions, in which the MZM splitting is finite and zero, respectively. The width in parameter space of the $\phi=\pi$ regions with MZMs pinned to zero is finite, unlike in topological Josephson junctions.
%This effect is reminiscent of the zero-energy pinning of finite-length topological wires with dielectric interactions.\cite{Dominguez:NQM17}However, here the CAR-induced zero-energy pinning is independent of the environment or of interaction strengths: it is of purely quantum origin, and not an effect of electronic incompressibility.

In essence, we have presented here a scheme towards one-dimensional topological superconductivity that extends  previous approaches that are based on the proximity effect, i.e. local Andreev reflections, of spinless helical electronic phases coupled to superconductors. While such approaches indeed produce a topological order parameter, its phase is fixed by the parent superconductor. In contrast, \emph{crossed} Andreev reflections, relevant in geometries as those discussed here, also produces a topological order parameter, but its sign may be either the same as or opposite to that of the parent, depending on the CAR amplitude itself. Controlling the sign of the topological gap in a stable way has many ramifications. We have shown how it may be exploited to produce stable, self-tuned $\pi$-junctions, wherein sizeable Majorana overlaps, which are problematic in more conventional Majorana devices, are no longer a concern, at least for pairs of MZMs at the junction. As a result, parametric braiding of Majoranas through e.g tunnel-braiding schemes becomes significantly more realistic.
The specific implementation of the CAR-induced topological gap described here is just one conceptually simple possibility, but it is not unique. Other phases, such as quantum anomalous Hall states, could also exhibit the requisite $\nu=1$ spin-singlet states.
%, trading the added complexity for less stringent conditions on spin-orbit coupling or Zeeman fields.
The temperature requirements for using our protocol are limited by both the Zeeman splitting and $\Delta^*$, which gives a conservative estimate between 0.1 K and 1 K, well within reach of current experiments on this type of systems. CAR-induced topological superconductivity is thus proposed as a promising road forward towards the next landmark in the field, the realisation of protected non-Abelian operations in the lab. 

%%%%%%%%%%%%%%%%%%%%%%%%%%%%%%%%%%%%%%%%%%%%%%%%%%%%%%%%%%%%%%%%%%%%%%%%%%%%%%%%

\appendix
\section{} \label{appA}
For the numerical calculations, we consider a two-dimensional square lattice that extends from $-L/2$ to $L/2$ along the $x$ axis, and from $-W/2$ to $W/2$ along the $y$ axis. The central superconducting strip, oriented along the $x$ axis, occupies the area that goes from $y = -W_{S}/2$ to $y = W_{S}/2$. We use either periodic boundary conditions (PBC) or open boundary conditions (OBC) along both directions, as specified in the main text. The tight binding Hamiltonian that we use for all the calculations in the paper is given by
\begin{widetext}
\begin{equation}
H = H_0 + H_Z + H_{S} + H_{SOC}
\end{equation}
where
\begin{equation}
H_0 = -\sum_{mn} \mu_{n} c_{mn}^{\dagger}c_{mn} - t \sum_{\langle m n,m'n'\rangle} c_{mn}^{\dagger}c_{m'n'} e^{-i\phi_{mn,m'n'}} \label{H0}
\end{equation}
\begin{equation}
H_Z = \sum_{mn} V_{n}^Z c_{mn}^{\dagger} \sigma_x c_{mn}
\end{equation}
\begin{equation}
H_{S} = \sum_{mn} \Delta_{n} \left[ c_{mn,\downarrow} c_{mn,\uparrow} + c_{mn,\uparrow}^{\dagger} c_{mn,\downarrow}^{\dagger}  \right]
\end{equation}
\begin{equation}
H_{SOC} = i\sum_{\langle m n,m'n'\rangle} (\alpha_{n} / a^2) \,\, c_{mn}^{\dagger} \left(\boldsymbol{\sigma}\times\textbf{r}_{mn}\right)_z c_{m'n'}
\end{equation}
%\end{widetext}
Where
\begin{itemize}
\item $\textbf{r}_{mn}=(ma, na)$, with $a$ the lattice parameter of the square lattice.

\item $\langle m n,m' n'\rangle$ indicates restriction to nearest neighboring sites.

\item $\mu_{n}=\mu_{N}$ for $n\,\in\,[-W/2,-W_{S}/2]$ and $n\,\in\,[W_{S}/2,W/2]$, and $\mu_{n}=\mu\neq\mu_{N}$ for $n\,\in\,[-W_{S}/2,W_{S}/2]$.

\item $\phi_{mn,m'n'}$ is the Peierls phase acquired by the electrons under an external magnetic field, defined as $\phi_{mn,m'n'}=\int_{r_{m'n'}}^{r_{mn}}\textbf{A}\cdot d\textbf{r}$ if $n\,\in\,[-W/2,-W_{S}/2]$ and $n\,\in\,[W_{S}/2,W/2]$ and that is 0 if $n\,\in\,[-W_{S}/2,W_{S}/2]$ due to the Meissner effect. Under the choice of the Gauge $\textbf{A}=(A_x(na),0,0)$, with
\begin{equation}
A_x(na) =
\left\lbrace
\begin{array}{ll}
B(na + W_{S}/2) & \mbox{ for } n\,\in\,[-W/2,-W_{S}/2]\\
0 & \mbox{ for } n\,\in\,[-W_{S}/2,W_{S}/2]\\
B(na - W_{S}/2) & \mbox{ for } n\,\in\,[W_{S}/2,W/2]
\end{array}
\right.
\end{equation}
and performing the integral, $\phi_{mn,m'n'}$ becomes
%$\phi_{mn}=Ba^2(m-m')(n+n')/2$.
\begin{equation}
\phi_{mn,m'n'} =
\left\lbrace
\begin{array}{ll}
Ba(m-m')\left[a(n+n')/2 + W_{S}/2\right] & \mbox{ for } n\,\in\,[-W/2,-W_{S}/2]\\
0 & \mbox{ for } n\,\in\,[-W_{S}/2,W_{S}/2]\\
Ba(m-m')\left[a(n+n')/2 - W_{S}/2\right] & \mbox{ for } n\,\in\,[W_{S}/2,W/2]
\end{array}
\right.
\end{equation}

\item $V_{n}^Z=V_Z\neq 0$ for $n\,\in\,[-W/2,-W_{S}/2]$ and $n\,\in\,[W_{S}/2,W/2]$ and $V_{n}^Z=0$ for $n\,\in\,[-W_{S}/2,W_{S}/2]$.

\item $\Delta_{n}=0$ for $n\,\in\,[-W/2,-W_{S}/2]$ and $n\,\in\,[W_{S}/2,W/2]$ and $\Delta_{n}=\Delta\neq 0$ for $n\,\in\,[-W_{S}/2,W_{S}/2]$.

\item $\alpha_{n}=\alpha\neq 0$ for $n\,\in\,[-W/2,W/2]$.

\item The creation and annihilation operators are two-component vectors in spin space
$$
c_{mn}^{\dagger}=\left(c^{\dagger}_{mn,\uparrow},c^{\dagger}_{mn,\downarrow}\right)
$$
\end{itemize}
\end{widetext}
The parameters used for the simulations that are common to all the results presented in the main text are $m^*=0.015m_e$, $B=0.34$ T, $\Delta=0.38$ meV, $\alpha = 3 \cdot 10^{-11}$ eV m, $V_{Z} = 0.3$ meV.
In addition, in Fig. \ref{fig:fig2}a we have employed a chemical potential of the proximitized region of $\mu=4$ meV for both panels, while changing the width from $W_S=300$ nm (left panel) to $W_S=2$ $\mu$m (right panel).
The chemical potential employed in Fig. \ref{fig:fig2}b is fixed to $\mu=0.74$ meV, while the width $W_S$ varies from $0$ to $950$ nm. The ratio $\mu/\Delta$ is therefore equal to 1.95. In Fig. \ref{fig:fig2}c we have used a strip 3 $\mu$m long with PBC. The blue points represent the lowest eigenvalues corresponding to a uniform chemical potential of $\mu_1=10.5$ throughout the strip, whereas the red ones represent the case of a strip that is cut in two halves, one with $\mu_1=10.5$ meV and the other with $\mu_2=13.1$ meV, characterized by gaps $\Delta^*$ of opposite sign. In Fig. \ref{fig:fig2}d we have used $W_S=220$ nm, $\mu_1=13.3$ meV and $\mu_2=16.5$ meV. The total length of the strip is of $3.4$ $\mu$m, and the length of the strip is of $2.4$ $\mu$m. The same width and chemical potentials have been used in Figures \ref{fig:fig3}b and \ref{fig:fig3}c, except for the fact that the strips are now spatially separated by 1 $\mu$m and long 2 $\mu$m each. We have considered a system with PBC and excluded the eigenvalues associated to the external MZMs (identically zero) that are present in the case of $\Delta^*_1\Delta^*_2<0$ (cfr Fig. \ref{fig:fig3}a). The hopping amplitudes from the MZMs to the dot are $t_1=0.32$ meV and $t_2=0.51$ meV.

%%%%%%%%%%%%%%%%%%%%%%%%%%%%%%%%%%%%%%%%%%%%%%%%%%%%%%%%%%%%%%%%%%%%%%%%%%%%%%%%

\section{} \label{appB}
To derive the analytical dependence of the gap $\Delta^*$ on the parameters of the system, we consider an infinite system with a superconducting strip coupled to its surrounding 2DEG by a real hopping $t'$ that could in principle be different from $t$ in Eq. \eqref{H0}. For $t'=0$ gapless edge states circulate along the 2DEG surface. A finite $t'$ couples edge states at either side of the strip, opening a gap $\Delta^*$ in their spectrum. This can be understood by considering the effective Hamiltonian of the 2DEG $H_\mathrm{eff} = H_0 + H_Z + H_{SOC} + \Sigma(\omega)$ once the strip is integrated out, which introduces a self-energy $\Sigma(\omega)$ that pairs opposite edge states. The induced superconducting pairing $\tilde\Delta$ is given by the off-diagonal (pairing) elements of the self-energy at $\omega=0$ between opposite edges (the actual gap $\Delta^*$ depends on this pairing $\tilde\Delta$, but also on the singlet amplitude of the 2DEG edge states, determined by $H_{SOC}$ and to be discussed later).\footnote{We may neglect the frequency dependence of $\Sigma(\omega)$ for the purpose of computing $\Delta^*$ as long as $\Delta^*\ll\Delta$, which is the physically relevant situation.} The self-energy from the strip reads
\begin{equation}
\Sigma(\omega) = \left.t'^*G^{(0)}_\mathrm{tb}(\omega;y,y')t'\right|_{y=0, y'=W_S}.
\label{TGT}
\end{equation}
Here $G^{(0)}_\mathrm{tb}$ is the tight-binding Nambu-Green function of the decoupled strip, evaluated above at $y, y'$ in opposite edges. A given $k_x$ wavevector is implicit here, as we assume $x$ translation symmetry. For simplicity we have shifted the strip to $y\,\in\,[0,W_{S}]$ here. In the continuum limit $a\to 0$ the Green's function is $G^{(0)} = \lim_{a\to 0} G^{(0)}_\mathrm{tb}/a$. We may decompose $G^{(0)}$ in terms of the continuum eigenvalues $\epsilon_\lambda$ and eigenvectors $\varphi_\lambda$ as %($\hbar=1$)
$$
G^{(0)}(\omega;y,y')=\sum_\lambda
\frac{\varphi_\lambda(y)\otimes \varphi_\lambda^\dagger(y')}{\omega-\epsilon_\lambda}
$$
This is a $4\times 4$ matrix, as $\varphi$ contains both spin and electron/hole amplitudes. The continuum Green's function, evaluated at the boundaries of the decoupled strip, vanishes by definition. One cannot, therefore, simply replace $G^{(0)}_\mathrm{tb}$ with $G^{(0)}$ in Eq. \eqref{TGT}.
%Instead, one can show that at $y=0,y'=W_{S}$ we have\cite{Prada:EPJB04}
As shown in [\onlinecite{Prada:EPJB04}], the Green's function at the outermost sites of a system described by a simple tight binding model can be written, in the limit where the lattice constant is the smallest length scale in the problem,  as:
$$
G^{(0)}_\mathrm{tb}(\omega;y,y') = - a^3\partial_y\partial_y'G^{(0)}(\omega,y,y')
$$
Hence, the pairing induced on the 2DEG edge state through crossed-Andreev reflection (CAR) processes reads
\begin{equation}
\tilde\Delta \approx -a^3 {t'}^2\left[\partial_{y}\partial_{y'}F^{(0)}(\omega=0;y,y')\right]_{y=0,y'=W_{S}}
\end{equation}
where $F^{(0)}=\frac{1}{4}\mathrm{Tr}(\tau_y\sigma_yG^{(0)})$ is the off-diagonal (pairing, or anomalous) component of the continuum Green's function and $\tau,\sigma$ are Pauli matrices in the particle-hole and spin sectors, respectively.

\begin{figure}
\centering
\includegraphics[width=8.7cm]{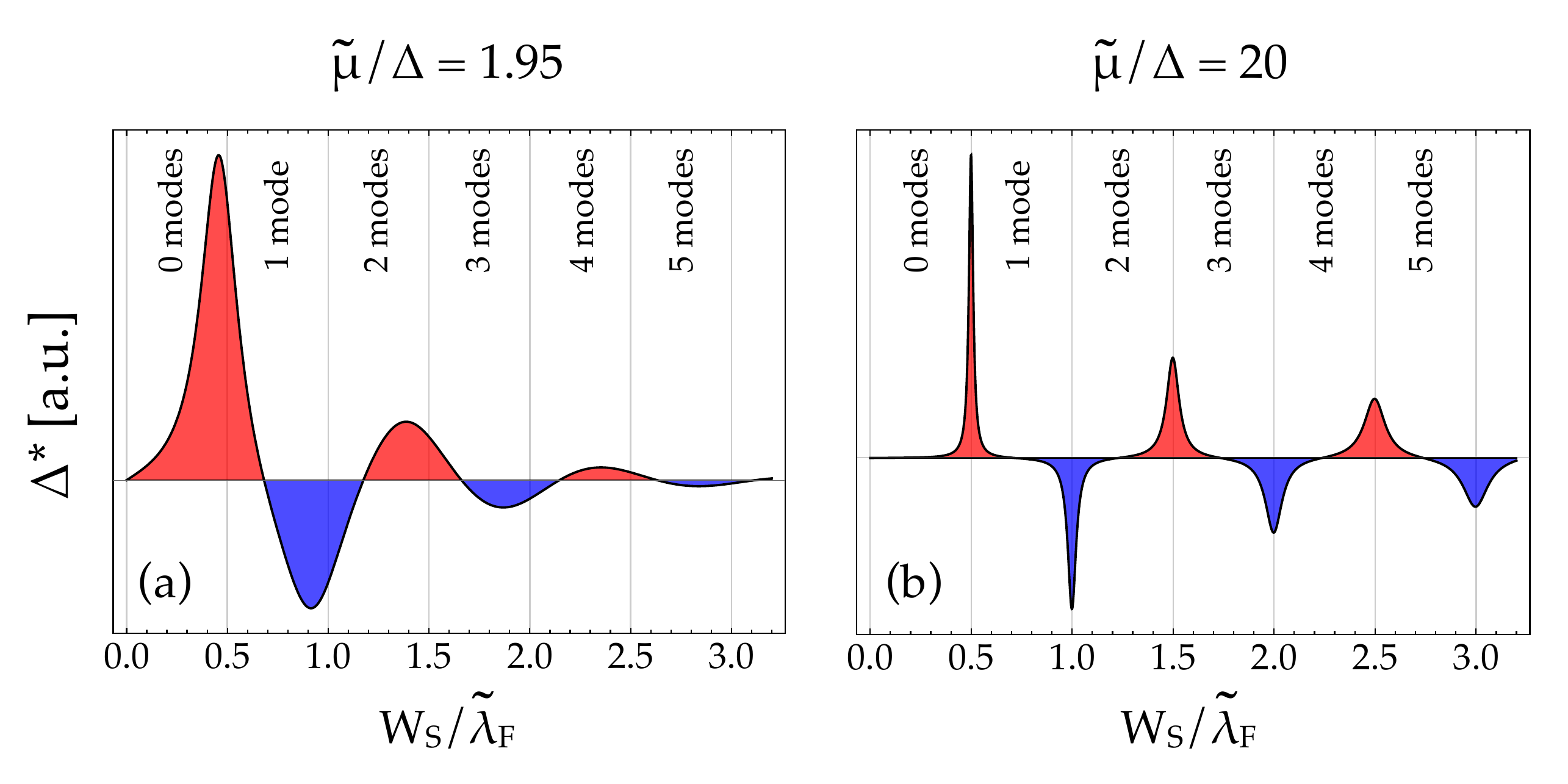}
\caption{Behaviour of the gap $\Delta^*$ obtained analytically as a function of $W_S/\tilde{\lambda}_F$. Vertical lines indicate the number of normal modes $\lfloor 2W_{S}/\tilde{\lambda}_F\rfloor$ in the strip. Panel (a) refers to a ratio $\tilde\mu / \Delta$ of the order of the unity (that matches value of the ratio $\mu/\Delta$ employed in the main text, cfr Fig. \ref{fig:fig2}b), whereas (b) refers to a ratio $\tilde\mu / \Delta$ that is one order of magnitude larger.}
%In the latter case the peaks are found at exactly half integer values of $W_S/\tilde{\lambda}_F$, while in the former they are slightly shifted with respect to these values.}
\label{fig:SI}
\end{figure}

To compute $G^{(0)}$ analytically we assume spin-orbit to be negligible inside the strip (it is assumed finite in the 2DEG only). Hence $G^{(0)}$ is spin degenerate, and can be obtained by diagonalising the $2\times 2$ continuum Hamiltonian of the strip
$$
H_{S}=\left(\frac{k_y^2}{2m}-\tilde{\mu}\right)\tau_z
+\Delta\tau_x,
\quad
\quad
\tilde{\mu}=\mu-\frac{k_x^2}{2m}
$$
where $m$ is the effective mass, $\mu$ is the chemical potential and $\Delta$ the pairing potential. The $\tau$ matrices are now Pauli matrices acting in a Cooper-pairing sector of a given spin, defined by the basis $\psi=\left(\psi_\uparrow,\psi_\downarrow^\dagger\right)^T$. The eigenvalues of this Hamiltonian are given by
$$
\epsilon_\eta=\eta\sqrt{\left(\frac{k_y^2}{2m}-\tilde{\mu}\right)^2 + \Delta^2}=\eta\sqrt{\xi^2 + \Delta^2}, \quad \eta=\pm 1
$$
where we have defined $ \xi=k_y^2/2m-\tilde{\mu}$. The associated normalized spinors are
$$
\varphi_\eta=
 \left(
\begin{array}{c}
u_\eta\\
v_\eta\\
\end{array}
\right)
=
\frac{1}{\sqrt{2\epsilon_\eta}}
 \left(
\begin{array}{c}
\eta\sqrt{\epsilon_\eta+\xi}\\
\sqrt{\epsilon_\eta-\xi}\\
\end{array}
\right)
$$
%where
%$$
%u_\eta= \sqrt{\frac{\epsilon_\eta+\xi}{2\epsilon_\eta}},
%$\quad\quad 
%v_\eta= \sqrt{\frac{\epsilon_\eta-\xi}{2\epsilon_\eta}}
%$$
For a given eigenvalue of the problem, the most general eigenstate solution is given by
$$
\varphi_\eta(y)=
 \left(
\begin{array}{c}
u_\eta\\
v_\eta\\
\end{array}
\right)
\left[
A_{\eta}e^{i k_y y} + B_{\eta}e^{-i k_y y}
\right]
$$
The coefficients $A_\eta$ and $B_\eta$ are found by imposing the boundary conditions that the wavefunction of the isolated strip needs to vanish at the boundaries:
$$
\varphi_\eta(y=0)=\varphi_\eta(y=W_{S}) = 0
$$
that yields $A_\eta=-B_\eta$ and the quantization of the wavevector along the $y$ direction
$$k_y^n=n\pi/W_{S}$$
The eigenvalues $\epsilon_\lambda$ and eigenvectors $\varphi_\lambda$ of the isolated strip, indexed by $\lambda=(n,\eta)$ quantum numbers, therefore read
$$
\epsilon_\eta^n=\eta\tilde\mu\sqrt{\left(\frac{n^2\tilde{\lambda}_F^2}{4W_{S}^2}-1\right)^2 + \left(\frac{\Delta}{\tilde{\mu}}\right)^2}
$$
and
$$
\varphi_\eta^n(y)=
%\left(
%\begin{array}{c}
%u_\eta^n\\
%v_\eta^n\\
%\end{array}
%\right)
\frac{1}{\sqrt{2\epsilon^n_\eta}}
 \left(
\begin{array}{c}
\eta\sqrt{\epsilon^n_\eta+\xi_n}\\
\sqrt{\epsilon^n_\eta-\xi_n}\\
\end{array}
\right)
\sqrt\frac{2}{W_S}\sin\left(\frac{n\pi y}{W_{S}}\right),
$$
where $\tilde{\lambda}_F=2\pi/\sqrt{2m\tilde\mu}$ and $\xi_n=\tilde \mu\left[(n\tilde{\lambda}_F/2W_{S})^2 -1\right]$. The Green's function of the isolated system is
$$
G^{(0)}(\omega;y,y')=\sum_{n=1}^\infty G_n^{(0)}(\omega;y,y')
$$
where
$$
G_n^{(0)}(\omega;y,y')=
\sum_{\eta=\pm 1}
\frac{\varphi_\eta^n(y) \otimes \left[\varphi_\eta^n(y')\right]^\dagger}{\omega-\epsilon_\eta^n}
$$
%Taking $\omega$ to zero
%$$
%G_n^{(0)}(\omega=0;y,y')=-\sum_{\eta=\pm 1}
%\frac{\varphi_\eta^n(y) \otimes \varphi_\eta^n(y')^*}{\epsilon_{\eta}^n}
%$$
The off-diagonal component of this matrix, $F_n^{(0)}=\frac{1}{2}\mathrm{Tr}(\tau_x G_n^{(0)})$, evaluated at $\omega=0$, reads
%$$
%G_n^{(0),o}(y,y')= \frac{-4 m^2 W_{S}^4 \Delta \sin(k_x^n y) \sin(k_x^n y')}{4 m^2 W_{S}^4 \Delta^2 + (n^2 \pi^2 - 2 m W_{S}^2 \tilde{\mu})^2}
%$$
$$
F_n^{(0)}(\omega=0; y,y')=\frac{2\sin(k_y^n y) \sin(k_y^n y')}{W_S}\sum_{\eta=\pm 1}\frac{\sqrt{(\epsilon_\eta^n)^2-\xi_n^2}}{2(\epsilon_\eta^n)^2}
%- \frac{\Delta\sin(k_x^n y) \sin(k_x^n y')}{\Delta^2 + \frac{(n^2 \pi^2 - 2 m W_{S}^2 \tilde{\mu})^2}{4 m^2 W_{S}^4}}
$$
The double derivative evaluated at the boundaries is
\begin{eqnarray*}
\left[\partial_{y}\partial_{y'}F^{(0)}_n(\omega=0;y,y')\right]_{y=0,y'=W_{S}}&& \\
= \frac{1}{W_{S}^3\Delta}\frac{(n\pi)^2 \cos(n\pi)}{1+(\tilde\mu/\Delta)^2\left[1-(n\tilde{\lambda}_F/2W_{S})^2\right]^2}&&
%\frac{4 \Delta  (n\pi m W_{S})^2\cos(n\pi)}{4 m^2 W_{S}^4 \Delta^2 + (n^2 \pi^2 - 2 m W_{S}^2 \tilde{\mu})^2}
\end{eqnarray*}
Performing the sum over $n$ we get that the effective pairing induced by the strip in the external edge states is
%$$
%\tilde{\Delta}=-(t'a)^2\sum_{n=1}^\infty
%\left[\partial_{y}\partial_{y'}F^{(0)}_n(\omega=0;y,y')\right]_{y=0,y'=W_{S}}=
%$$
\begin{eqnarray}
\tilde{\Delta}&=&-a^3 {t'}^2\sum_{n=1}^\infty
\left[\partial_{y}\partial_{y'}F^{(0)}_n(\omega=0;y,y')\right]_{y=0,y'=W_{S}}\nonumber \\
&=&
a^3 {t'}^2\frac{1}{\Delta W_{S}^3}\frac{4\pi^2}{(\tilde{\lambda}_F/W_{S})^2\tilde\mu/\Delta} \label{Deltatilde}\\
&&\times\,\mathrm{Im}\left[\frac{2\pi\sqrt{i+\tilde\mu/\Delta}}{\sqrt{\tilde\mu/\Delta}\,\tilde{\lambda}_F/W_{S}}\csc\left(\frac{2\pi \sqrt{i+\tilde\mu/\Delta}}{\sqrt{\tilde\mu/\Delta}\,\tilde{\lambda}_F/W_{S}}\right)\right].\nonumber 
\end{eqnarray}
%\redmark{A related expression was found in Ref. \onlinecite{Reeg:PRB17} in the particular limit of large $W_{S}$.}
Related expressions were derived in Ref. 24 corresponding to various limiting cases of the general result above.
Recall that all the dependence on the $k_x$ wavevector is inside $\tilde\mu=\mu-k_x^2/2m$ and $\tilde\lambda_F=2\pi/\sqrt{2m\tilde\mu}$. At $k_x=0$ these quantities become the actual Fermi energy $\mu$ and Fermi wavelength $\lambda_F=2\pi/\sqrt{2m\mu}$ of the superconducting strip, respectively. The ratio $\lfloor 2W_{S}/\tilde{\lambda}_F\rfloor$ represents the total number of open modes in the quasi-1D strip with a given $k_x$  in the absence of superconductivity. Equation \ref{Deltatilde} then shows that the \emph{sign} of the CAR-induced pairing is given by the parity of the number of open modes, see Fig. \ref{fig:SI}.
%We note that the sign of the gap is periodic when expressed as a function of $W_{S}/\tilde{\lambda}_F$, but not if plotted versus the more natural $W_{S}/\lambda_F$ for $k_x=k_F^N$ (see below) as done in the main text.

%where we recall that 
%$\tilde{\mu}=\mu-k_x^2/2m$. The relevant value of $k_x$ corresponds to the Fermi wavevector of the edge states in the 2DEG, so that $\tilde\mu \approx\mu-(k^\mathrm{QH}_F)^2/2m = \mu-\mu_{QH}$, where $k^\mathrm{QH}_F$ and $\mu_{QH}$ denote the Fermi wavevector and Fermi energy of the non-proximitized 2DEG region, respectively.

%We notice that for this derivation we didn't take into account that the proximized region is provided with SOC as well. This is a good approximation in the limit where $\epsilon_{SOC}=m\alpha^2 / 2 \hbar^2 \ll \mu$. 

We now consider how the presence of spin-orbit coupling in the 2DEG allows for the pairing $\tilde \Delta$ to open a gap $\Delta^*$ in the edge state spectrum. 
%on the SOC is entirely accounted for by the dependence on the SOC of the edge states of the external 2DEG. 
We employ a simplified low-energy description of the edge states. Given that the 2DEG bulk is insulating, we consider just the 1D chiral channels generated at the two sides of the strip in the $\nu=1$ QH regime. These can be modelled by the 4x4 continuum Hamiltonian
\begin{equation}\label{OregLutchyn}
H=\left(\frac{k^2}{2m}-\mu_N+V_{Z}\sigma_x +\alpha k\sigma_y\right)\tau_z -\tilde{\Delta}\tau_y\sigma_y \nonumber
\end{equation}
where $k=k_x$, $\mu_N$ is the chemical potential, $\alpha$ is the spin-orbit coupling and $\tilde{\Delta}$ is the CAR-induced pairing derived above (evaluated at $k_x=k_F$, i.e. at the wavevector for which the edge states cross zero energy). We recall that the $\sigma$ matrices act in spin space and the $\tau$ matrices in particle/hole space. This Hamiltonian is akin to the Oreg-Lutchyn model\cite{Lutchyn:PRL10,Oreg:PRL10}, and is a valid description of the edge states at either side of the strip at low energies. For Zeeman fields $V_Z<\mu_N$ the model has two carrier species propagating along each direction, which corresponds to filling $\nu=2$ of the QH state. The model is then in a topologically trivial phase. For strong enough Zeeman fields $V_Z>\sqrt{\tilde{\Delta}^2+\mu_N^2}$, however, it can be driven into a topologically non-trivial phase. One spin sector is thus depleted, so that the corresponding filling is $\nu=1$ in the absence of the superconducting strip (one mode propagating along each direction on each side of the strip). The pairing $\tilde\Delta$ then creates a topologically non-trivial gap $\Delta^*$ that leads to Majorana bound states.

The wavefunction satisfying the Schroedinger equation is now a 4-component spinor $\psi=(\psi_\uparrow,\psi_\downarrow,\psi_\uparrow^\dagger,\psi_\downarrow^\dagger)^T$. At $\mu_{N}=0$, the system opens a gap of
$$
\Delta^*(k)= \frac{\sqrt{k^4 + 4 m (m \gamma_k - \Gamma_k)}}{m}
$$
where
$$\gamma_k = V_Z^2 +  \alpha^2 k^2+ \tilde{\Delta}^2$$
and
$$\Gamma_k=\sqrt{ \alpha^2 k^6 + V_Z^2 k^4 + 4 m^2 V_Z^2 \tilde\Delta^2}$$
If we work within the limit in which $\tilde\Delta$ is the smallest scale of the problem, then the wavevector at which the gap opens is well approximated by the Fermi wavevector at zeroth order in $\tilde\Delta$. Thus
$$
k=k_F \approx \sqrt{2 m^2 \alpha^2 + \sqrt{m^2 (V_Z^2 + m^2 \alpha^4)}}
$$
and, therefore,
$$
\Delta^*=\Delta^*(k_F)= 2 \sqrt{\beta\left[1 + \frac{\tilde{\Delta}^2}{\beta} - \sqrt{1 + 
 \tilde{\Delta} \frac{4 V_Z^2}{\beta^2}}\right]}
$$
where
$$
\beta = 2 V_Z^2 + 4 m^2 \alpha^4 + 
  4 m \alpha^2 \sqrt{V_Z^2 + m^2 \alpha^4}=
  2 (V_Z^2 + \alpha^2 k_F^2)
$$
Now one can expand $\Delta^*$ in series as a function of $\tilde{\Delta}$ up to first order, obtaining
$$
\Delta^*\approx
 \sqrt{1 - \frac{2 V_Z^2}{\beta}} \tilde{\Delta}=
 \frac{\alpha k_F}{\sqrt{V_Z^2 + \alpha^2 k_F^2}} \tilde{\Delta}
$$
By writing the Zeeman and Rashba part of the Hamiltonian as
$$
H_{Z+SOC}=\textbf{h}\cdot\boldsymbol{\sigma}
$$
where $\textbf{h}=(V_Z, \alpha k_F,0)$. This expression allows to define a canting angle $\theta$ such that
$$
\theta=\arcsin\left(\frac{\alpha k_F}{\sqrt{V_Z^2 + \alpha^2 k_F^2}}\right)
$$
and
$$
\Delta^*=\sin\theta\tilde{\Delta}
$$
This angle represents the spin-orbit-induced deviation of the edge state spins away from the Zeeman axis $\sigma_x$. Now, plugging in the values that we used in the main text for the numerical calculations, we obtain the behaviour of $\Delta^*$ as a function of $W_S/\tilde{\lambda}_F$ shown in Fig. \ref{fig:SI}a, in excellent agreement with the full numerics shown in the main text. (Note that in the main text $W_S$ is normalized to $\lambda_F$ in spite of the $\tilde\lambda_F$ used in Fig. \ref{fig:SI}).

\section{} \label{appC}
In this section, we study the stability of the $\pi$-junction to perturbations in the phase difference between $\Delta^*_1$ and $\Delta^*_2$. To confirm that a $\pi$-junction in $\Delta^*$ is indeed a stable solution for the system, one must demonstrate that a phase difference $\phi^*=\pi$ between $\Delta^*_1$ and $\Delta^*_2$ corresponds to a minimum in the Josephson free energy under variations of $\phi^*$. Bardeen et al.\cite{Bardeen:PR69} and Beenakker and van Houten\cite{Kirk:AP92} demonstrated, using complementary approaches, that the free energy of a generic Josephson junction may be written, at finite temperature $T$ and up to a phase-independent constant, as
\begin{equation}
E_J(\phi)= -k_B T \sum_{\epsilon_n<0}\ln\left( 2 \cosh \frac{\epsilon_n(\phi)}{2k_B T}\right)
\end{equation}
where the sum is performed over both spin flavours and over both particle- and hole-like levels.  In the low temperature limit $E_J(\phi)$ reduces to
\begin{equation}
\lim_{T \rightarrow 0}E_J(\phi)= - \frac{1}{2} \sum_{\epsilon_n<0}\epsilon_n(\phi)
\end{equation}
Variations in the phase of the induced gap $\phi^*$ can be generated through variations in the phase $\phi$ of the left and right portions of the parent superconductor on top of the two half-strips, which is a controllable parameter in the model. Hence, computing the free energy $E_J(\phi)$ and knowing the relation between $\phi^*$ and $\phi$, one may establish whether the $\pi$-junction is stable. Intuitively the total free energy, which includes the energy of the parent superconductor, will be a competition between the phase rigidity of $\phi$ and the phase rigidity of $\phi^*$, which in the $\pi$-junction configuration are out of phase.

\begin{figure}
\centering
\includegraphics[width=\columnwidth]{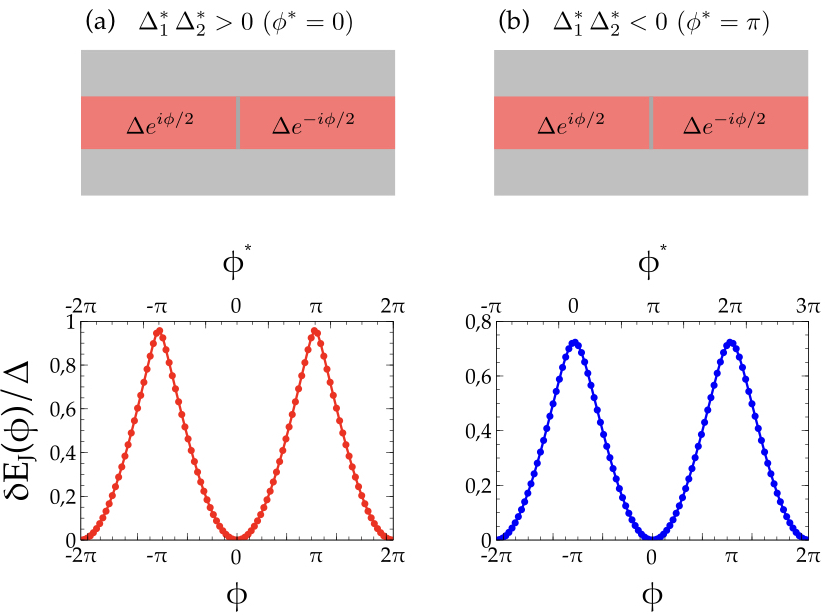}
\caption{Two junctions (a) and (b) differ in $W_S\lambda_F$, so that (a) is a conventional Josephson junction and (b) is a topological $\pi$ junction with $\Delta^*_1\Delta^*_2<0$. (c,d) The total free energy for (a,b), respectively, as a function of a phase difference $\phi$ in the parent superconductor, or $\phi^*$ in the induced gap $\Delta^*_{1,2}$. The minimum of the free energy shows that the junctions are thermodynamically stable at $\phi^*=0$ in (a) and $\phi^*=\pi$ in (b).}
\label{fig:R5}
\end{figure}

We have performed numerical calculations of the Josephson free energy of the junction as a function of parent phase difference $\phi$ $\in$ $[- 2 \pi: 2 \pi]$ in two cases. In case (a) the ratio $W_S/\lambda_F$ on the two sides of the junction is such that for $\phi=0$ the induced gap has also $\phi^*=0$ (panel a of Fig. \ref{fig:R5}), and the $\pi$-junction case (b) which has $\phi^*=\pi$ at $\phi=0$ (panel b of Fig. \ref{fig:R5}). The results for the free energy (panels c,d, respectively) show that the minimum is obtained in both configurations when $\phi=0$. In other words, the phase rigidity of the parent superconductor is stronger than that of the induced gap, and hence determines the equilibrium configuration. It is more expensive to generate a phase difference of $\pi$ in $\phi$ than in $\phi^*$. Thus, in case (b), the $\phi^*=\pi$ configuration is thermodynamically stable.

\acknowledgements

We are grateful to L. Chirolli, E. Prada, C. Reeg, J. Klinovaja and D. Loss for fruitful discussions. F. F. and F. G. acknowledge the financial support by Marie-Curie-ITN Grant No. 607904-SPINOGRAPH. F. F. and P.S-J. acknowledge financial support from the Spanish Ministry of Economy and Competitiveness through Grant No. FIS2015-65706-P (MINECO/FEDER).

\bibliography{biblio}

\end{document}